\pdfoutput=1
\documentclass{article}

\usepackage{arxiv}

\usepackage[utf8]{inputenc} 
\usepackage[T1]{fontenc}    
\usepackage{url}            
\usepackage{booktabs}       
\usepackage{amsfonts}       
\usepackage{nicefrac}       
\usepackage{microtype}      
\usepackage{lipsum}
\usepackage{graphicx}
\usepackage{makecell}
\usepackage{graphicx}
\usepackage{subcaption}
\usepackage{natbib}
\usepackage{listings}
\usepackage{hyperref}
\hypersetup{colorlinks}

\title{Master of Disaster: A Disaster-Related Event Monitoring System From News Streams}

\author{
 Junbo Huang \\
  Department of Computer Science\\
  University of Hamburg\\
  Vogt-Kölln-Straße 30, 22527 Hamburg, Germany\\
  \texttt{junbo.huang@uni-hamburg.de} \\
   \And
 Ricardo Usbeck \\
  Head of the AI and Explainability Group\\
  Leuphana University Lüneburg\\
  Universitätsallee 1, 21335 Lüneburg, Germany \\
  \texttt{ricardo.usbeck@leuphana.de} \\
}

\begin{document}
\maketitle
\begin{abstract}
  The need for a disaster-related event monitoring system has arisen due to the societal and economic impact caused by the increasing number of severe disaster events. An event monitoring system should be able to extract event-related information from texts, and discriminates event instances. We demonstrate our open-source event monitoring system, namely, Master of Disaster (MoD), which receives news streams, extracts event information, links extracted information to a knowledge graph (KG), in this case Wikidata, and discriminates event instances visually. The goal of event visualization is to group event mentions referring to the same real-world event instance so that event instance discrimination can be achieved by visual screening.
\end{abstract}

\section{Introduction}
The increased frequency and magnitude of disasters in recent years have led to significant human, environmental, and economic losses. In response to this, disaster-related event monitoring systems, such as GDELT~\citep{leetaru2013gdelt} and GDACS~\citep{de2006global}, have emerged as critical tools to (1) monitor events and (2) provide timely overviews for an early disaster response planning. Such large-scale systems combine manual and automated approaches in retrieving event data from sensory and raw text data. Advances in machine learning have significantly boosted predictive models’ ability to extract accurate information from raw texts in the above-mentioned monitoring systems. However, a major challenge on discriminating event instances remains unsolved. In other words, how do we know if multiple news articles are referring to the same real-world event instance? 

Event instance discrimination is a less research area. Existing work mainly relies on entity matching based on factors such as location and time~\citep{DBLP:conf/semweb/RolloP20, DBLP:conf/ijcai/ZavarellaPITA20, DBLP:conf/icsai/AiXSWM21}, or linking event mentions to entries in a knowledge graph (KG) like Wikidata\footnote{\url{https://www.wikidata.org/}} ~\citep{DBLP:conf/eacl/YuYGR23}. However, heuristic-driven methods often suffer from poor generalization, while KG-driven approaches are limited by the inability to represent events absent from the KG. In this word, we strive for an event monitoring system that (1) receives news stream; (2) extracts event information; (3)links extracted information to a Knowledge Graph (KG), in this case, Wikidata and (4) discriminates event instances visually.

To achieve this goal, we suggest that a half-automated, human-in-the-loop event monitoring system can leverage predictive models to provide reliable information for humans. Such information includes users, geospatial and temporal information of events as well as an indication on whether two or more event mentions are referring to the same real-world event instance.

In this work, we present Master of Disaster (MoD)\footnote{The online demo can be found here \url{https://hitec.demo.skynet.CoyPu.org/}. Please ignore tabs other than Event Extraction and Event Visualization. The source code for the GUI can be found here \url{https://github.com/semantic-systems/SEMS-tool-suite/}. The source code for the event extractor can be found here \url{https://github.com/semantic-systems/CoyPu-EventExtraction}}, a half-automated, human-in-the-loop disaster-related event monitoring system with three key features. First, MoD receives daily news streams from GDELT\footnote{\url{https://www.gdeltproject.org/}} with a keyword search. Second, MoD extracts event-related information with a transformers-based event extractor, whose output is linked to Wikidata and the CoyPu ontology\footnote{\url{https://schema.coypu.org/global/}}. Third, MoD provides a visualization of events to represent the similarity of different news in the embedding space. 
To compensate for the error produced by the event extractor, a visualization from an unsupervised clustering algorithm is provided as a reference for the human evaluation of event distributions.

MoD aims to provide an overview of daily events extracted from the user-provided keywords. Furthermore, it should be mentioned that the semantic output (linking to Wikidata) helps leverage downstream tasks such as knowledge validation and common sense reasoning.

\section{Master of Disaster (MoD)}
\begin{figure}
\centering
\includegraphics[width=0.8\textwidth]{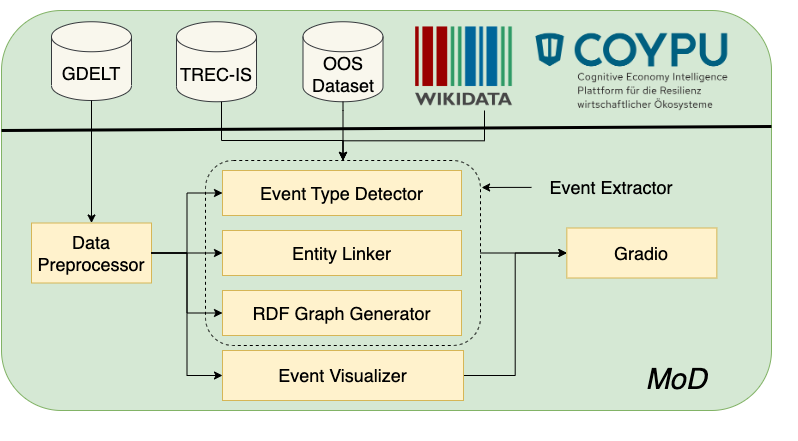}
\caption{Overview over MoD's Architecture.}\label{mod}
\end{figure}

MoD is an open-sourced application with four components: a data preprocessor, an event extractor, an event visualizer, and a gradio-based GUI\footnote{\url{https://gradio.app/}}. 
The architecture of MoD is illustrated in Figure~\ref{mod}.

\subsection{Data preprocessor}
First, the data preprocessor queries a list of articles containing the user-provided keywords with the GDELT API\footnote{Note that the GDELT API restricts the maximal amount of returned articles to 250.}. Furthermore, it applies a filtering method to select articles that are written in English. This is because the event extractor can only receive English inputs, currently. With the filtered articles, the data preprocessor takes the title of each article to represent an event. Additionally, all artifacts (such as white spaces around punctuation marks) are removed from the titles to retrieve the final representation of events in the textual space. 

\subsection{Event Extractor}
The event extractor consists of three components, an event type detector, an entity linker and an RDF graph generator\footnote{Resource Description Framework (RDF) is a standard model for data interchange on the Web.}.
The \textbf{event type detector} is a pre-trained \textit{RoBERTa-base}~\citep{DBLP:journals/corr/abs-1907-11692} model fine-tuned on the TREC-IS dataset~\citep{DBLP:conf/iscram/McCreadieBS19}. The model is optimized using the supervised contrastive learning~\citep{DBLP:conf/nips/KhoslaTWSTIMLK20} loss. 
In the original TREC-IS benchmark, the task was to predict information types from raw texts, such as \textit{Request-GoodsServices} or \textit{Report-EmergingThreats}. However, we modified the original task schema to predict the event type\footnote{In total, there are 9 event types: \textit{tropical storm}, \textit{flood}, \textit{shooting}, \textit{covid}, \textit{earthquake}, \textit{hostage}, \textit{fire}, \textit{wildfire} and \textit{explosion}} (such as \textit{Earthquake} or \textit{Shooting}). Note, the event type was considered a meta-information in the original dataset. Also, the dataset did not include an \textit{out-of-scope (oos)} class describing non-events. Thus, we included data from the TweetEval dataset~\citep{DBLP:conf/emnlp/BarbieriCAN20} to represent an \textit{oos} class. It is worth noting that TweetEval included 7 subtasks, such as \textit{Emotion Recognition}, \textit{Irony Detection} and \textit{Hate Speech Detection}. Sentences in TweetEval are all user-generated tweets, which contain a variety of topics, language styles and structures than traditional news articles. By integrating such samples in the training data, the model learned a more concentrated feature representation for the given event classes.

Next, we employed an \textbf{entity linker} to extract event-related entities. 
MoD presently considers two relations, namely \textit{hasLocality} and \textit{hasImpactOn}, following the CoyPu Event ontology\footnote{\url{https://schema.CoyPu.org/global/2.3\#Event}}. 
Here, we leveraged a local entity linker, namely BLINK\footnote{\url{https://github.com/facebookresearch/BLINK}}~\citep{DBLP:conf/emnlp/WuPJRZ20}, a \textit{BERT}-based entity linking model which links entities to Wikipedia. 
However, since the CoyPu ontology is based on Wikidata as background KG, the \textbf{RDF graph generator} maps from Wikipedia to Wikidata and outputs an event graph in JSON-LD (see Appendix \ref{jsonls}). 

\subsection{Event Visualizer}
\begin{figure}[ht!]
\begin{minipage}{0.48\textwidth}
    \includegraphics[width=\linewidth]{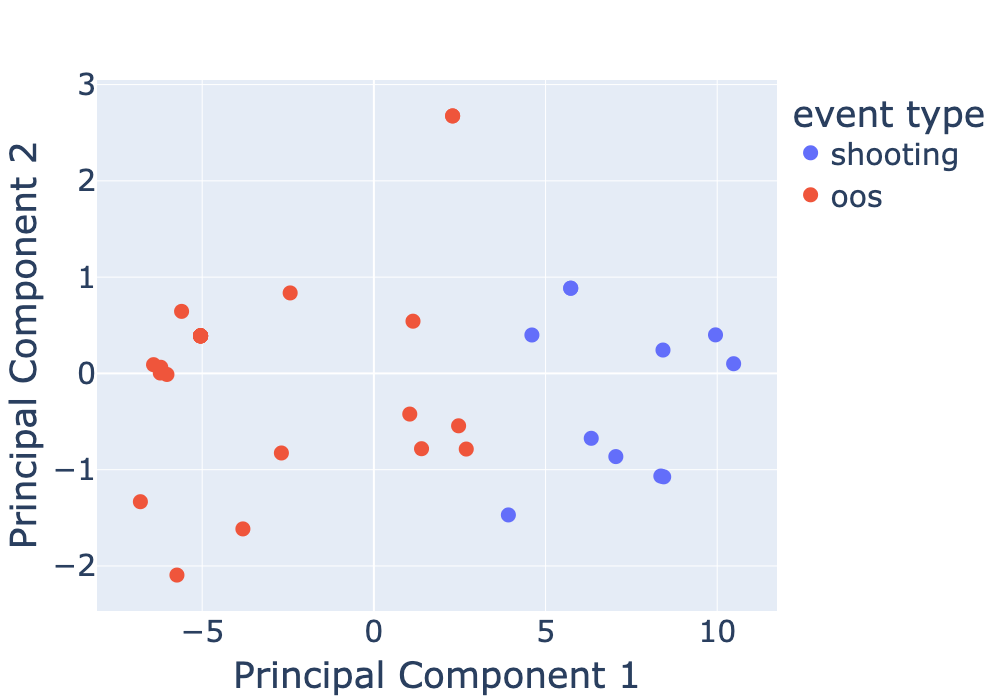}
\end{minipage}
\begin{minipage}{0.49\textwidth}
    \includegraphics[width=\linewidth]{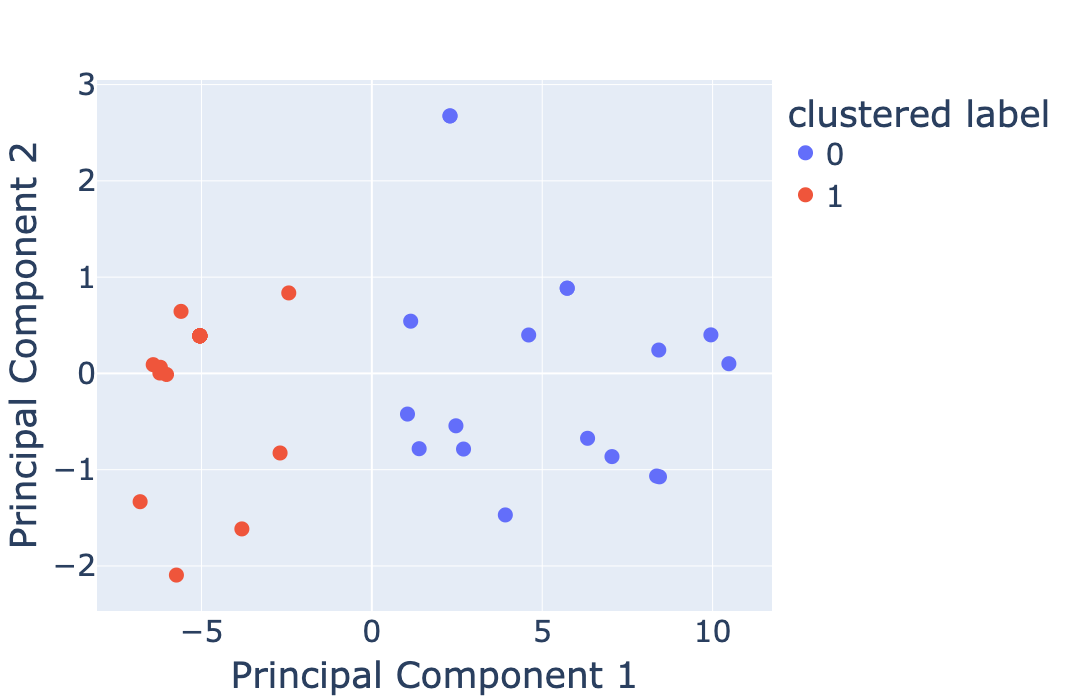}
\end{minipage}
\caption{Event Visualization with respect to the classification (left) and clustering (right) Result. The search keyword is "Hamburg", and the date is "11. March 2023"}
\label{visualization}
\end{figure}

The event visualizer (1) receives a stream of filtered articles from the data preprocessor, (2) followed by the event type detector to generate sentence embeddings from the last hidden layer and make predictions on event types, and (3) visualizes the sentence embeddings in a two-dimensional coordinate. The goal of event visualizer is to group event mentions referring to the same real-world event instance. We applied a dimensionality reduction technique based on Principal Components Analysis (PCA). Two resulting two-dimensional visualizations are presented in MoD, as shown in Figure~\ref{visualization}. 

The first one illustrates the event type prediction by the event type detector (left-hand side in Figure \ref{visualization}) and the second visualization (right-hand side in Figure \ref{visualization}) introduces the prediction from a clustering algorithm. We used DBSCAN~\citep{DBLP:conf/kdd/EsterKSX96} as the clustering algorithm because DBSCAN is a density-based algorithm, where the number of clusters is not pre-defined. Therefore it can group news into an arbitrary number of event instances. From the visualization shown in Figure~\ref{visualization}, we have observed that DBSCAN considers local structures (measured by the distance between sentence representations) of sentence embedding in the PCA-reduced two-dimensional space. This provides insights for human evaluators to have a direct access to event distribution, which is important to discriminate between event instances.

The final component in MoD is the \textbf{gradio-based interactive GUI} which is a centralized, injective component that makes API calls to all services used in MoD. These services include an event extractor and an event visualizer. 

\section{Evaluation}

To ensure that MoD can be practically deployed as a half-automated disaster-related event monitoring system, we conducted a survey with seven NLP researchers, among which five of the participants are frequent news readers. The only demographic that we controlled for is the expertise in NLP. We observed that six participants have expertise in knowledge graphs and three participants have expertise in event extraction. 
The main investigation in the survey is the usability of MoD, which consists of the following questions on a 5-point Likert scale. 
\begin{enumerate}
    \item How easy is it to use the Event Extraction tab? (1: very difficult; 5: very easy)
    \item How accurate is the information provided by the event extractor? (1: very inaccurate; 5: very accurate)
    \item How easy is it to use the Event Visualization tab? (1: very difficult; 5: very easy)
    \item Is the overview of daily news by the Event Visualizer informative? (1: very non-informative; 5: very informative)
    \item Will you use Event Visualizer for fast access to the news? (1: no; 5: yes)
\end{enumerate}
As a result, we received 4.14 points on average on both the first and the third question (level of easiness to use the system); 3 points on average on the third question (accuracy by the event extractor); 3.57 points on average on the fourth question (informativeness of the event visualizer) and 3.14 point on average on the last question (practicality). Albeit with a small set of participants, we observed that the majority of the participants agreed that MoD is easy to use and the visualization is informative. 

\section{Conclusion and future work}
In this paper, we presented MoD, a disaster-related event monitoring system. Moreover, we explained the architecture of MoD and the functionality of each component. 
The goal of MoD is to provide an event monitoring system that (1) reveives news stream; (2) extracts event information; (3) links the extracted information to a KG and (4) discriminates event instances mentioned in the news. 
In future work, we aim to provide a temporal analysis of events and increase the performance of the event extractor.

\section*{Ethics Statement}
To our knowledge, this work does not concern any substantial ethical issue. Corpora used in this work are preprocessed by masking all user mentions and links. Of course, the application of classification algorithms could always play a role in Dual-Use scenarios. However, we consider our work as not risk-increasing.

\section*{Acknowledgements}
The authors acknowledge the financial support by the Federal Ministry for Economic Affairs and Energy of Germany in the project CoyPu (project number 01MK21007[A-L]). HITeC is “G”
This work was also supported by the House of Computing and Data Science (HCDS) of the Hamburg University within the Cross-Disciplinary Lab programme, and by the Ministry of Research and Education within the project ‘RESCUE-MATE: Dynamische Lageerstellung und Unterstützung für Rettungskräfte in komplexen Krisensituationen mittels Datenfusion und intelligenten Drohnenschwärmen’ (FKZ 13N16844).

\bibliographystyle{unsrtnat}  
\newpage
\bibliography{references}

\appendix

\section{Example output graph of the event extractor}
\begin{lstlisting}
{
@id:  "https://data.CoyPu.org/event/mod/d868a8be-c49e-48b8-a3a5-5b4b12d7d97f",
@type:  [
    0:  "https://schema.CoyPu.org/global#Event" ,
    1:  "http://www.wikidata.org/entity/Q2252077"
],
http://www.w3.org/2000/01/rdf-schema#comment:  [
    0:  {
        @value:  "Hamburg shooting : Multiple dead after attack at 
        Jehovah Witness church in Germany"
    }
],
https://schema.CoyPu.org/global#hasImpactOn:  [
    0:  {
        @id:  "http://www.wikidata.org/entity/Q35269"
    }
],
https://schema.CoyPu.org/global#hasLocality:  [
    0:  {
        @id:  "http://www.wikidata.org/entity/Q1055"
    } ,
    1:  {
        @id:  "http://www.wikidata.org/entity/Q183"
    }
],
https://schema.CoyPu.org/global#hasPublisher:  [
    0:  {
        @value:  "HiTec"
    }
],
https://schema.CoyPu.org/global#hasTimestamp:  [
    0:  {
        @value:  "15_03_2023_17_57_56"
        }
    ]
}

\end{lstlisting}
\label{jsonls}
\end{document}